\documentclass[10pt]{article}

\usepackage{infocomp}
\usepackage{graphicx}
\usepackage{setspace}

\newcommand{\ProportionOfText}{0.48}

\usepackage{fancyhdr}
\vspace{-10cm}
\abstract{
In the last years, Distributed Visualization over Personal Computer (PC) clusters has become important for research and industrial communities. They have made large-scale visualizations practical and more accessible. In this work we survey Distributed Visualization techniques aiming at compiling last decade's literature on the use of PC clusters as suitable alternatives to high-end workstations. We review the topic by defining basic concepts, enumerating system requirements and implementation challenges, and presenting up-to-date methodologies. Our work fulfills the needs of newcomers and seasoned professionals as an introductory compilation at the same time that it can help experienced personnel by organizing ideas.
}

\address{Universidade Federal de S\~ao Carlos at Sorocaba, Universidade Federal do ABC*,\\ Universidade de S\~ao Paulo at S\~ao Carlos+\\
$[$junio,lzaina$]$@ufscar.br,
andre.balan@ufabc.edu.br,
agma@icmc.usp.br}

\title{A Survey on Distributed Visualization Techniques over Clusters of Personal Computers}

\author{
Jose Rodrigues, Andre Balan*, Luciana Zaina, Agma Traina+
}

\keywords{Distributed visualization, scientific visualization, computer graphics, distributed computing}

\vspace{-4cm}
\receivedate{July 16, 2009}

\acceptdate{October 16, 2009}

\begin{document}

\maketitle

\section{Introduction}
\label{Introduction}


\noindent{Visualizing huge datasets demands processing power that surpasses commonplace workstations. To cope with this need, high-end systems using Symmetrical Multi-Processors technology (SMP) \cite{96} have been commercialized. Alternatively, during the last decade, a great number of works have discussed matters of design and implementation of visualization over computer clusters. These works constitute the Distributed Visualization discipline and range from complete systems to prototypes and theories to better utilize distributed computation.}

According to the Top 500 Supercomputer list \cite{98}, currently more than 40 percent of the fastest computers in the world are clusters of networked computers. Among these clusters are the Quake project \cite{82} at the Pittsburgh Supercomputing Center, which is composed of hundreds of proprietary systems in a cluster of workstations reaching top generation performance. 

\begin{figure}[ht]
		\centering		
		\includegraphics[width=\ProportionOfText\textwidth]{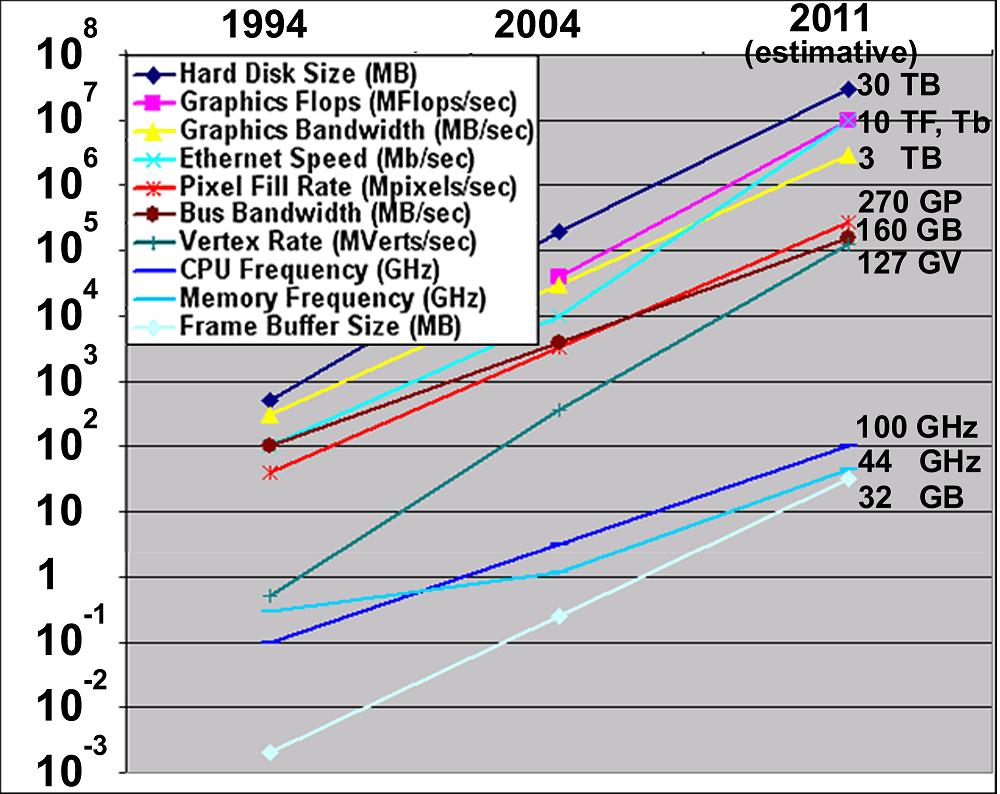}		
		\caption{Various aspects of commodity hardware evolution and tendency for a single processing node equipped with a top generation Graphics Processing Unit. Except for the network speed data, all the data were extracted from Fernando's seminar \cite{93}.}
	  \label{ComodityEvolution}		
\end{figure}

\indent{Concomitant to this process, the PC commodity industry has furiously evolved in the last years tending to continue at this pace, see figure \ref{ComodityEvolution}. Advances in storage devices, processing power, memory speed, graphical buses and frame buffers have doubled every period around two years or less. These resources lead to a graphical power that was formerly infeasible. Today, a commodity U\$1500 workstation has graphics capabilities that exceed those of a late 1990s U\$500K supercomputer. These advances, together with network improvements, made it possible to build PC clusters to rival high-end machines.}

\indent{An up-to-date benchmark \cite{99} from the National Aeronautics and Space Administration agency (NASA) demonstrates not only that commodity computer clusters rival to SGI workstations, but also that the maintenance cost of these workstations is sufficient to build a new PC cluster every year. Thus, in this work, we compile the progresses in Distributed Visualization focusing on clusters of PCs. We build a condensed document aiming at organizing ideas and discussing unsolved issues.}

\indent{In this text we review the complexities to consider when designing and implementing Distributed Visualization over clusters of PCs. At the same time, we present the advantages of this platform, including low costs for supercomputing, the ability to track technologies, systems that can be incrementally upgraded, open source software and vendor autonomy.}

\indent{The organization of the text is as follows. Section \ref{Section_Distributed_Visualization} introduces the Distributed Visualization topic and section \ref{Section_Implementation_Issues} presents a set of issues related to implementing clusters of PCs. Section \ref{Section_Distributed_Parallel} presents libraries that intend to abstract the assembling of such systems and section \ref{Section_Conclusions} concludes the paper.}

\section{The Visualization Pipeline}
\label{Section_Distributed_Visualization}

\begin{figure}[ht]
		\centering		
		\includegraphics[width=\ProportionOfText\textwidth]{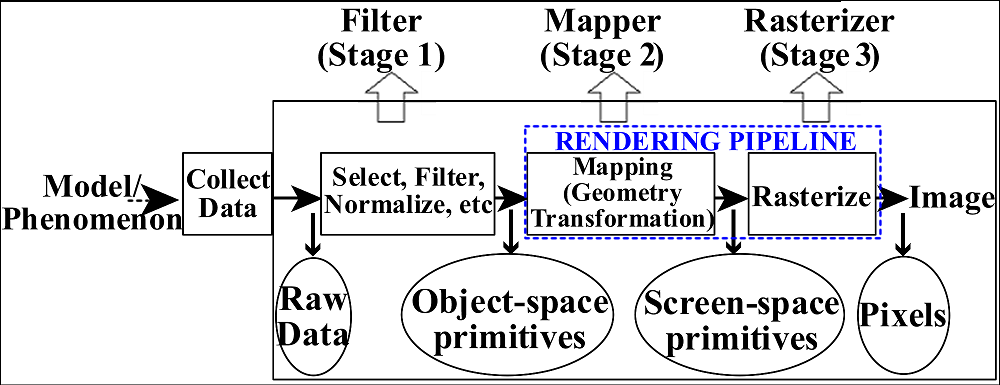}		
		\caption{Three stages named filter, mapper and rasterizer compose the complete visualization pipeline. The first of them may be omitted, executed before or along the visualization process. In such case, only the mapper and the rasterizer stages compose the process. These two stages determine the core of the visualization pipeline, which is called {\it rendering pipeline} (dashed rectangle).}
	  \label{Pipeline}		
\end{figure}

\noindent{Visualization is organized according to the model cast by two works, Upson {\it et al} \cite{71} and Haber and McNabb \cite{32}. Their model, presented in figure \ref{Pipeline}, describes three stages to achieve a visualization. In a pipeline structure, each stage executes a distinct processing whose output feeds the next stage.}

\begin{figure}[ht]
		\centering		
		\includegraphics[width=\ProportionOfText\textwidth]{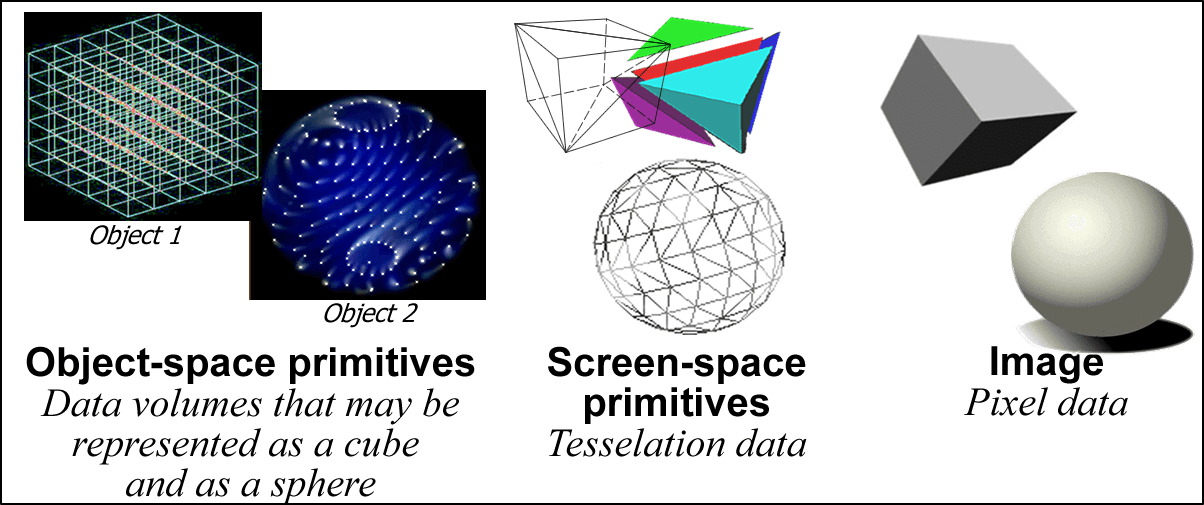}		
		\caption{Illustration of data that may take part of a visualization pipeline. Two objects, a cube and a sphere, their correspondent volumes and tesselation data followed by renderization.}
	  \label{fig:PipelineData}		
\end{figure}

\indent{After an initial data collecting stage, the pipeline proceeds with the raw data volume that is processed according to the purposes of the intended visualization. In this stage, named filter (preprocessing or traversal), the dataset is selected, filtered, cleaned, enriched, summarized, normalized, and/or submitted to any useful processing to optimize the rendering process, e.g., culling operations. Then, as illustrated in figure \ref{fig:PipelineData}, the prepared data (object space primitives) are submitted to a mapping procedure (geometry transformation) to determine how the data will be displayed in the form of geometrical entities (screen-space primitives), e.g., tesselation data, an isosurface or a contour map. The last stage, named rasterizer (or renderer), applies operations like projection, lightening and/or shading to the screen-space primitives in order to finally generate the image (pixels). The whole process is known as {\it visualization pipeline}, while the last two stages are known as {\it rendering pipeline}. We widely refer to the above concepts along the text.}

In the Distributed Visualization domain, the
stages of the visualization pipeline are not restricted to a single machine or location. Each step may take place locally or remotely, at the client that will analyze the data or at the server that owns the processing power. Accordingly, we define Distributed Visualization as the use of distributed resources to drive visual analysis. It is expected that such systems present the possibility of disjoining data and exploration sites, the possibility of combining autonomous processing resources and the possibility of collaborative work.

\newpage
Advances in commodity hardware have led to the use of PC clusters as an attractive option for Distributed Visualization. To explore these prospects, algorithms have been proposed to deal with memory and communication constraints in distributed environments. This approach increases the limits of a single PC, but it has to tackle with load balancing and inter-process communication, among other problems, that we discuss in this text.

\section{Design and Implementation Issues for Distributed Parallel Rendering Systems}
\label{Section_Implementation_Issues}

\noindent{Distributed Visualization systems have been oriented to PC clusters (distributed memory) architectures that concern parallel rendering in distributed systems. Motivations include: the low cost of commodity PC hardware is far smaller than that of high-end visualization systems; PCs are all-purpose machinery and can also be used for non-graphical applications; it is possible to benefit from the standards of the PC market, which allows for a continuous upgrade with reduced effort; the open-source movement provides high quality software at low costs; and, it is possible to add more PCs to the system in order to bear with power increase demands.}

The main consideration for Distributed Visualization over PC clusters is the algorithm that binds the distributed resources into the {\em visualization pipeline} defined in section \ref{Section_Distributed_Visualization}. Such algorithms fall into one of three categories: sort-first, sort-middle and sort-last (subsection \ref{Subsection_Algorithms}), having as their design goal concerns on load balancing and communication constrains among the nodes of the cluster (subsection \ref{Subsection_Load_Balancing}). Supporting this structure lies network technology (subsection \ref{Subsection_Network_Issues}), techniques for data management (subsectoin \ref{subsec:Data_Management}) and techniques for parallel storage (subsection \ref{subsec:Parallel_IO}). Another concern is the operating system over which the distributed visualization ensemble will execute (subsection \ref{Operating_System}). Alternatively, distributed parallel rendering libraries offer high-level implementation, as reviewed in the next section.\\

\subsection{Algorithms for parallel tiled Distributed Visualization}
\label{Subsection_Algorithms}

\noindent{In reference to the foundational visualization pipeline described in seciton \ref{Section_Distributed_Visualization}, Molnar {\it et al} \cite{50} formulate the most accepted classification for distributed parallel rendering algorithms. Their analysis determines how the visualization pipeline (geometric transformation followed by rasterization) maps onto a general parallel algorithm. According to the theory, the parallelism is achieved through the assigning of transformation tasks (on object-space primitives) and rasterization tasks (on image-space primitives) to the distributed processing units.}

\begin{figure}[ht]
		\centering		
		\includegraphics[width=\ProportionOfText\textwidth]{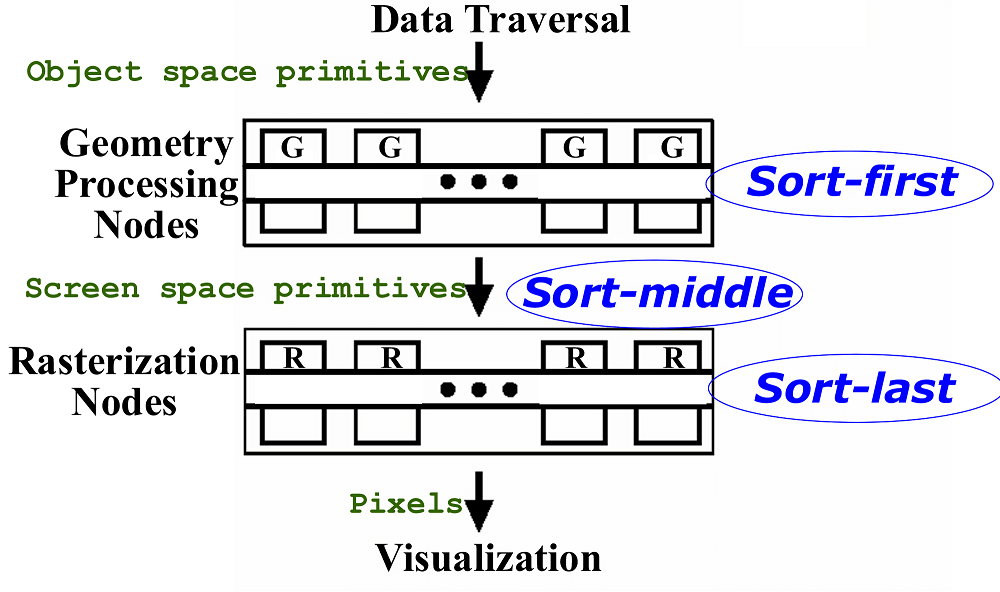}		
		\caption{Algorithms for distributed rendering rely on the decision of where/when the parallel tasks will be designated for the processing units. This is considered a sorting/classification problem. There are three possibilities for this sorting operation, each of which deeply characterizes the correspondent algorithms.}
	  \label{SortProblemMoments}		
\end{figure}

\indent{The idea is that the complex modeling of visualization scenes, geometrical entities (object-space primitives and screen-space primitives) and rendered pixels can be distributed in different ways across the processing resources. The design task, thus, worries about how to assign the data portions of the pipeline so that the intended image can be achieved at the end. At the same time, processing load and network communication constraints must be satisfied. Sutherland {\it et al} \cite{68} consider it a sorting (or classification) problem. This sorting can occur in one of three moments, as presented in figure \ref{SortProblemMoments}. After the sorting is complete, the data need to be redistributed among the nodes to reflect the new-sorted arrangement.}

\indent{The classes defined when considering the rendering pipeline and the sorting problem are named sort-first, sort-middle and sort-last. They differ by terms of bandwidth requirements, amount of duplicated work and load balance.}

\vspace{3mm}
\noindent\textbf{Sort-first (image (or pixel) space parallelism)}

\begin{figure}[ht]
		\centering		
		\includegraphics[width=\ProportionOfText\textwidth]{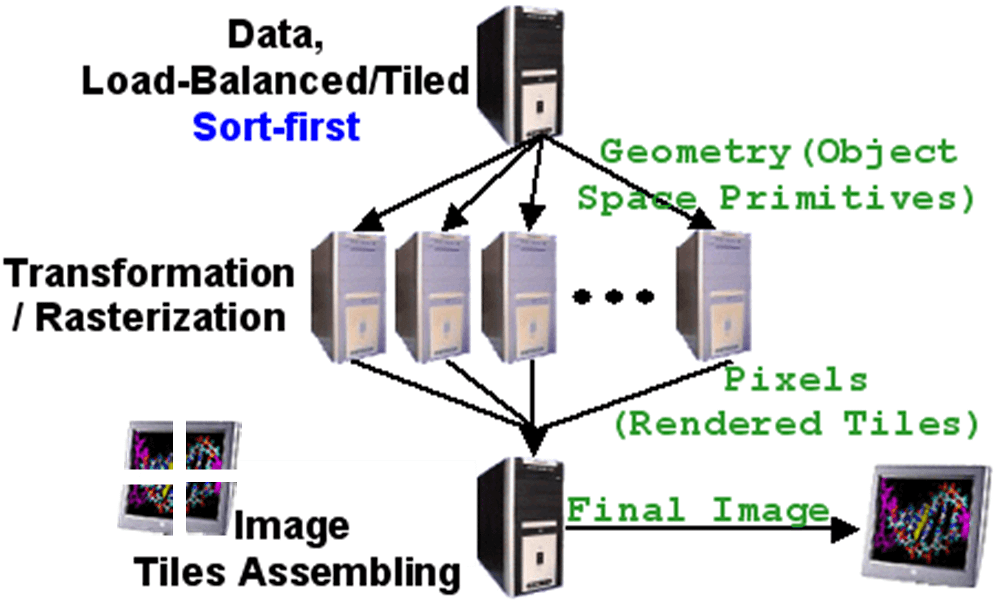}		
		\caption{Classical sort-first configuration. Based on tiled partitioning information, the data is distributed among the processing nodes. After transformation and rasterization, a final step assemblies the tiles to form the image.}
	  \label{Sort-first}		
\end{figure}

\noindent{In sort-first, the screen is divided into disjoint regions (tiles) that are assigned to the processing nodes, as illustrated in figure \ref{Sort-first}. To do so, the objects primitives are submitted to a minimum geometry transformation (screen-space bounding box calculation) necessary to determine which tile of the screen they overlap. Then, the objects primitives are transmitted to the processing nodes (renderers) that correspond to the tiles of the screen, where both transformation and rasterization are performed. Finally, the rendered tiles are reassembled to form the final scene.}

\indent{The initial geometry transformation phase of sort-first is called pre-transformation. Due to this extra processing, sort-first is the most expensive design for distributed visualization, but also the least bandwidth consumer. Examples include the work by Zhu {\it et al} \cite{77} and the work by Mueller \cite{54}. The later presents a study about sort-first implementation and its advantages, mainly the frame-to-frame coherence (high for interaction sequences) and the lower bandwidth demand.}\\

\noindent\textbf{Sort-middle}

\begin{figure}[ht]
		\centering		
		\includegraphics[width=\ProportionOfText\textwidth]{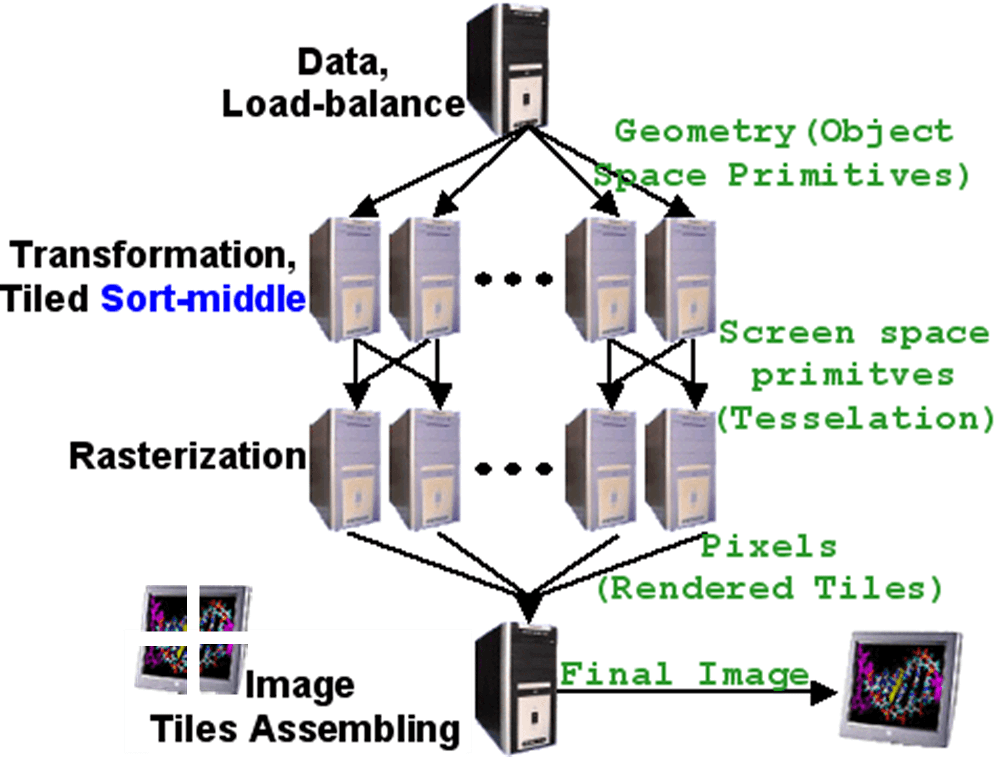}		
		\caption{Sort-middle configuration with two levels of processing nodes. The screen-space data communication may also occur in between nodes at the same level, determining a one-level-only structure, where transformation nodes also act as rasterization nodes.}
	  \label{Sort-middle}		
\end{figure}

\noindent{It is the natural approach for distributed parallel rendering because transformation and rendering are performed at different levels of the cluster, see figure \ref{Sort-middle}. Initially, the algorithm distributes object space primitives among the nodes according to some load balance method, e.g, round robin. Then, after geometry processing, the resultant screen space primitives are distributed to the rasterization nodes. Similar to sort-first, the algorithm assigns tiles of the screen to specific processing nodes but, differently, there is no pre-transformation step.}

\indent{The disadvantages occur when the tessellation ratio is high. Tessellation refers to the decomposing of larger primitives into smaller ones. It determines that the system must redistribute several display primitives instead of just one object primitive. For sort-middle, high tessellation ratios imply in higher communication costs.}

\indent{Another disadvantage of sort-middle is the load imbalance on the rasterization units if the primitives are unevenly distributed over the screen, what also may occur to sort-first algorithms. Also, according to Mueller \cite{54}, the loose coupling in the middle of the pipeline can limit feedback from the rasterization stage back to the transformation stage, what makes certain visibility culling algorithms either less efficient or infeasible. Montrym {\it et al} \cite{51} present a custom-designed implementation of this parallelism and Ellsworth \cite{20} makes an extensive review of sort-middle systems.}\\

\noindent\textbf{Sort-last (object-space parallelism)}

\begin{figure}[ht]
		\centering		
		\includegraphics[width=\ProportionOfText\textwidth]{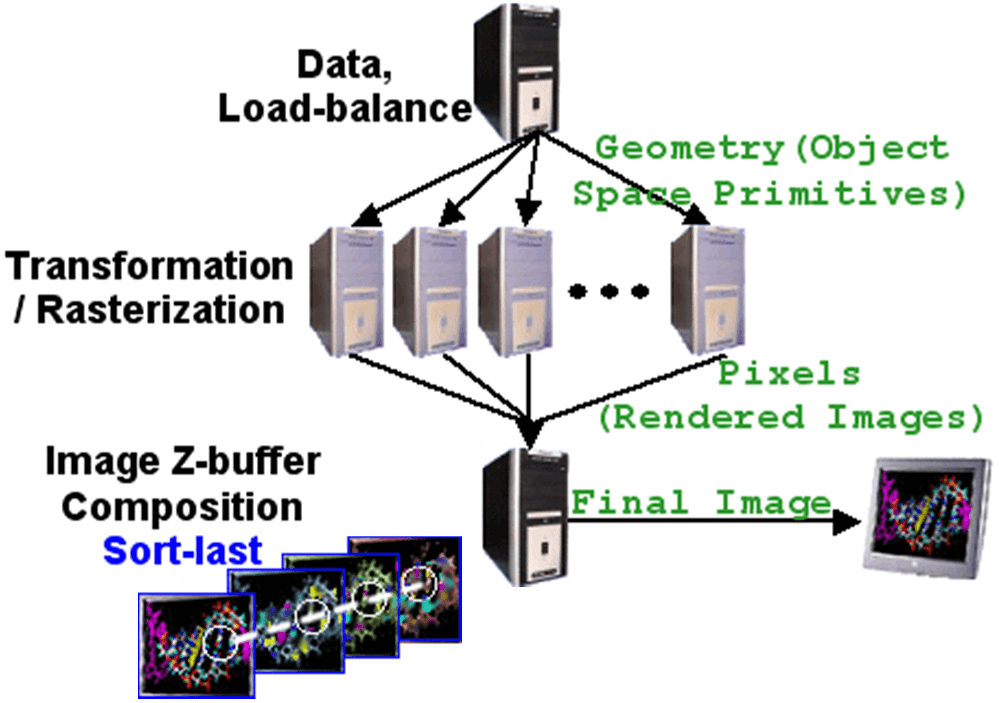}		
		\caption{Sort-last-full configuration. The initial distribution of tasks considers only load-balancing directives. In a second step, each processing node is responsible for the complete rendering of a full screen image with only a subset of the objects that compose the visualization. At the final step, each screen is superimposed following depth information in order to have the correct visibility of the visual entities.}
	  \label{Sort-last}		
\end{figure}

\noindent{In this case, the sorting occurs after the end of the rendering pipeline, that is, the pixels are ready to compose the image, as presented in figure \ref{Sort-last}. In a load-balanced manner, the processing nodes (renderers) receive arbitrary subsets of the object-space primitives. After transformation and rasterization, the resultant pixels are submitted to a composition procedure. At this final step, sort-last will have produced a set of full-screen images (sort-last-full) or a set of screen-space primitives (sort-last-sparse). These images are recomposed by hardware or software that compute every sample at each pixel to define the primitives' visibility. This is called (depth) sorting and, according to Foley {\it et al} \cite{25}, it relies on the use of Z-buffering. Thus, the processing nodes must send, along with the pixels, the correspondent Z-buffers. This need highly increases the required bandwidth to the order of gigabytes per second.}

\indent{The advantages of sort-last are the better control of load balance concerned to the object-space primitives and the simplicity of the approach because the processing nodes perform the full pipeline independently. Implementation examples include the works by Lombeyda {\it et al} \cite{44} and by Morel {\it et al} \cite{53}.}

\indent{In the literature, each author advocates for her/his choice considering the most suitable features for sort-first, sort-middle or sort-last. More recent works point to sort-first and sort-last to be used with PC cluster implementations. Sort-middle is used with high-end shared-memory systems as SGI's hardware, probably because fast memory buses are less influenced by the overheads of sort-middle. For comparison, in table \ref{TableAlgorithms} we present an overview of the three possibilities.}

\indent{Hybrid approaches are also possible, as done by Samanta {\it et al} \cite{64} and Garcia and Shen \cite{29}. The former work tries to minimize the sort-last composition overhead by means of a dynamic sort-first partition. Their approach benefits from a view-dependent partitioning of both the 3D model and the 2D screen. The later work leverages the advantages of both sort-first and sort-last approaches with a hybrid-sorting of both image and data partitioning for load balance.}


\begin{table*}[htp]
	\caption{Overview of the main algorithm possibilities for Distributed Visualization.}
	\label{TableAlgorithms}
\centering		
\begin{tabular}{llll}
		\hline
		  \textbf{Feature}& 
			\textbf{Sort-first}&
			\textbf{Sort-middle}&
			\textbf{Sort-last}\\
		\hline
		\hline			
			Parallelism&
			 image&
			 object / image&
			 object\\
		\hline
			 Sorted data&
			 object space primit.&
			 screen space primit.&
			 pixels (z-buffer)\\			
		\hline			
			Bandwidth demand&
			 low&
			 medium&
			 high\\			
		\hline
			 Overhead factor&
			 pre-transform./overlap&
			 high overlap&
			 image composition\\
		\hline
			 Frame coherence&
			 yes&
			 no&
			 no\\
		\hline
\end{tabular}
\end{table*}
\vspace{-0.2cm}

\subsection{Load balancing}
\label{Subsection_Load_Balancing}

\noindent{Load balance applies distinctly for image parallelism (sort-first, sort-middle) and object parallelism (sort-last). In object parallelism, object distribution-rules define how to reach nearly equal loads among the processing nodes. In image parallelism, load balance is based on screen partitioning methods.}

\newpage
Ellsworth \cite{20} points that, for object parallelism, random or round robin approaches are used to distribute objects among the processing nodes. These techniques work fairly well for objects with homogeneous size and complexity. For objects with great differences, the time to process them may vary by a large factor. Further possibilities for load balancing consider the geometry as hierarchical structures or as sets of primitives (flat structures), this topic is reviewed by Ellsworth {\it et al} \cite{111}.

\begin{figure}[ht]
		\centering		
\includegraphics[width=\ProportionOfText\textwidth]{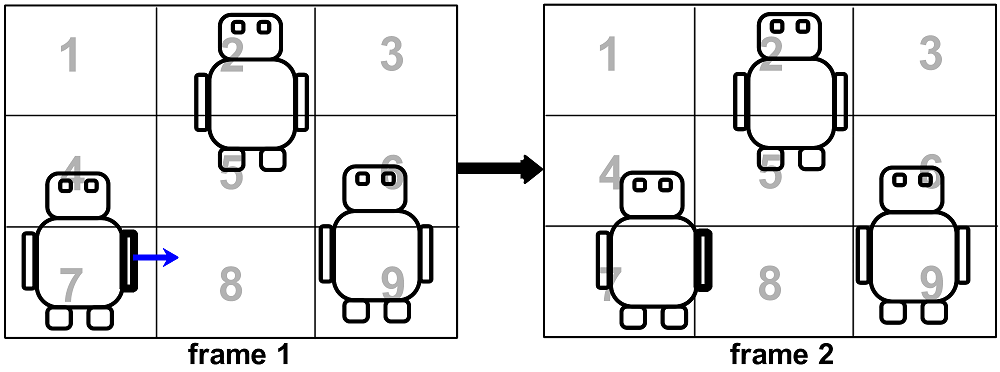}		
		\caption{Screen-partitioning with 9 tiles. By means of frame coherence, only the primitive highlighted in tile 7 must be sent in order to draw tile 8 in frame 2, which was formerly empty. The other primitives remain in memory via retained mode operation.}
	  \label{Screen_Partitioning_Frame_Coherence}		
\end{figure}

For image parallelism, if not equal portions of the image are assigned to the processing nodes, some of them will remain idle while waiting for others to finish their task. This problem is treated by screen dividing methods, as exemplified in figure \ref{Screen_Partitioning_Frame_Coherence}, which can be static or dynamic.

\indent{Static screen dividing methods divide the screen into more regions than the number of processing nodes and assign the regions in an interlaced fashion to these nodes. The number of regions per node is called granularity ratio. For low granularity the workload may not be balanced. For high granularity, we may have a high overlap factor (the average number of regions overlapped by the primitives), what leads to network overload in image parallelism. If a primitive lies over three regions, the entire primitive must be transmitted three times for transformation and rasterization because, even if just a small piece overlaps a tile, its computation depends on the entire primitive. According to Molnar {\it et al} \cite{50}, if we assume equal sized primitives and equal probability for the positioning of these primitives on the screen, the overlap factor is given by:}

\vspace{-0.4cm}
\begin{eqnarray}
Overlap = ( (R_{weight} + P_{weight} ) / R_{weigth} )* \\ \nonumber
( (R_{height} + P_{height} 
)/R_{height} )
\end{eqnarray}
\setlength{\arraycolsep}{5pt}
\vspace{-0.4cm}

Where $R_{weight}$, $R_{height}$, $P_{weight}$ and $P_{height}$ are, respectively, the weight and the height of a given screen region and the weight and the height of a given primitive bounding box.

Dynamic (adaptive) screen dividing methods work by computing statistics from on-screen primitives distribution in order to intelligently determine and assign the tiles. These algorithms add overhead to the visualization process first due to the gathering of statistics and decision making they implement; and second due to the more elaborated screen division. Two common approaches for adaptive screen division are: first to statically partition the screen and then dynamically allocate them to the processing nodes; another method is to first settle a constant number and assignment of regions and then dynamically vary their shape, as done by Roble \cite{61}. Also, dynamic partitioning combined with dynamic assignment is possible, as proposed by Mueller \cite{54}. Finally, a comparison of algorithms for space division is available in the work by Kurc {\it et al} \cite{42}.

\subsection{Network issues}
\label{Subsection_Network_Issues}

\noindent{The great disadvantage of PC clusters if compared to shared bus (SMP) systems is that data sharing does not occur over a high speed direct access memory bus, but over a much slower network. Thus, network factors demand special attention in order to reach effective bandwidth and network latency performances. Bandwidth corresponds to the available data/time transmission. Network latency is the time to prepare and transmit the data between two nodes.}

Bandwidth constraints are affected by the network speed, the data-network adapter speed, the bus interface and the memory. That is, the bandwidth is a function of the data traveling time, receiving time, in-node transfer time and memory storage time. 

Meanwhile, the network latency varies with the network interface that sends/receives the data, the bus interface to in-node read/transfer the data, the memory architecture to access/store the data and the processing power available to decide and perform the whole process. High network latency times barely affects scarce long message communication, e.g. Internet browsing, but it is decisive for communication characterized by plentiful short messages, as required by computer clusters.

These factors must be designed to maximize the bandwidth at the same time that the latency be minimized. Together with high quality hardware and system architecture, an appropriate network must be settled. A suitable practice is to isolate the cluster in a network in which traffic is limited to the cluster communication. This setting constitutes a System Area Network (SAN). In such systems, the hubs must have minimal retransfer latency and the interconnection of different networks must be avoided due to higher latency. For optimization, a switch device, instead of a hub, can serve the network so that intelligent directional ports permit parallel communication. Following we review major factors to come up with a suitable network structure for distributed parallel rendering.\\

\noindent\textbf{The Message-Passing Interface (MPI) standard}

\noindent{The MPI standard \cite{85}, is a message-passing library being developed since 1992. A message passing library is a high-level abstraction that permits inter-communication within a collection of autonomous processes each of which with its own local memory. It eases the implementation of shared memory and distributed shared memory systems, which are the foundation of computer clusters parallelism. There are several message-passing libraries, but the MPI standard is the {\it de facto} convention for clusters.}

MPI ranges from supercomputing to PC clusters. It is a standard and not a product, implementations of it are available for several operating systems and network technologies. Liu {\it et al} \cite{90} present a performance comparison of MPI implementations over InfiniBand, Myrinet and Quadrics technologies (detailed further in this section). Gropp {\it et al} \cite{132} describe the MPICH, a portable implementation of MPI (``CH'' stands for ``Chameleon''). MPICH is the most used free distribution of MPI and its design goal is to combine performance with high portability within a single implementation. Another popular implementation of MPI is Romio \cite{135}, with broad portability and free on-line support.\\

\newpage
\noindent\textbf{Via technology}

\noindent{In stacked-based communication protocols, as TCP/IP, great part of the latency is caused by the in-node operation. When a process in a cluster node wants to transmit, it prepares the data and makes a high-level call to some network API that access the network hardware. Then, when the respective interruption request (IRQ) is issued, the OS copies the data to some other memory space used as buffer. This buffer, finally, is read by the network hardware that transmits it via the physical layer. This process consumes a large amount of time per data transmission, lowering the process by up to orders of magnitude.}

To lessen this problem, it was conceived the Virtual Interface Architecture (VIA) \cite{129}. VIA, created by Intel, Compaq and Microsoft, describes an alternative interface between network hardware and user software. This interface provides direct access to the network hardware (user level communication protocol), what lowers the transmission latency by avoiding the OS intermediation (zero-copy protocol). VIA is accomplished by hardware integration on the Network Interface Card (native implementation), or by software emulation. The former achieves the best performance, the later consumes extra processing, but even though its latency performance surpass that of regular network usage, according to a IEEE report \cite{127}.

Cameron and Regnier \cite{129} present complete information about VIA. Baker {\it et al} \cite{130} describe a study on VIA performance gains over Gigabit Ethernet. In \cite{131} it can be found information about the MVICH (MPICH for Virtual Interface Architecture), a popular implementation of MPI on top of VIA technology. Also in \cite{131} it is described the Modular VIA (M-VIA) a high performance implementation of VIA for Linux systems. The use of VIA is a straight method to diminish the effects of network latency, which is the main limitation in computer clusters.\\

\noindent\textbf{Network technologies}

\noindent{Transmitting 3D objects over the network, as for sort-middle algorithms, can compromise the available bandwidth. Thus, research is performed to devise geometry compression algorithms, as done by Touma \cite{12}. These algorithms reduce bandwidth requirements to up to 10 bits per vertex, including connectivity. However, the required bandwidth and network latency still can overscale commodity Ethernet networks. To cope with that, high-performance interconnection technologies are used. We list them on the next paragraph.}

Myrinet \cite{88} is a packet-communication and switching technology used to interconnect clusters of workstations, it offers up to 2 Gigabit/second full duplex links and it is based on the ANSI (American National Standards Institute) Standard ANSI/VITA 26-1998. The Quadrics technology \cite{89} reaches up to 8.5 Gigabit/second full duplex rates on top hardware systems provided with the QsNet II network. InfiniBand \cite{91} is steered by an association of member companies involved in performance computing, data center and storage implementations. It offers up to 30 Gigabit/second channels. The Scalable Coherent Interface (SCI) network, which is an ANSI/ISO/IEEE Standard (1596-1992), has been specifically designed to computer clusters. With reduced network latency time, SCI behaves like a bus or a network using point-to-point links to achieve higher speed. It implements a cache scheme as a coherent virtual shared memory. Its Dolphin \cite{104} release reaches up to 2.6 Gigabit/second rates.

In table \ref{TableNetworks} we present an overview of these technologies together with the Gigabit Ethernet commodity technology. The data are only for rough comparison because a number of other factors may influence the performance and costs.

\begin{table*}[htp]
	\caption{Network technologies overview.}
	\label{TableNetworks}
\centering	
\begin{tabular}{lllll}
		\hline
		  \textbf{Network technology}& 
			\textbf{Bandwidth} \textbf{(MB/s)}&
			\textbf{Latency ($\mu$s)}&
			\textbf{Avg. Price/Port (U\$) \cite{110}}\\
		\hline
		\hline			
			Gigabit Ethernet&
			 $<$ 125&
			 $<$ 100&
			 $\sim$ 300.00\\
		\hline			
			10 Gigabit Ethernet&
			 $<$ 1250&
			 $<$ 60&
			 $\sim$ 7000.00\\			
		\hline
			 Myrinet \cite{88}&
			 $<$ 250&
			 $<$ 10&
			 $\sim$ 400.00\\
		\hline
			 Quadrics QsNet II \cite{89}&
			 $<$ 1064&
			 $<$ 3&
			 $\sim$ 2000.00\\
		\hline
			 Infiniband \cite{91}&
			 $<$ 3750&
			 $<$ 7&
			 $\sim$ 800.00\\
		\hline
			 SCI \cite{104}&
			 $<$ 326&
			 1-2&
			 $\sim$ 800.00\\
		\hline
\end{tabular}
\end{table*}

All these technologies have support for VIA (native or emulated) and for implementations of MPI. The choice for one of them depends on other factors such as compatibility with the cluster equipment and operating system, performance and price. Latency is decisive for massive short message communication, thus, the ill latency performance of Ethernet makes it the worst choice. Quadrics and Infiniband present superior bandwidth coupled with very attractive latency times. The drawback is the elevated price of these options. More adequate alternatives are Myrinet and SCI networks. Myrinet has already been widely used for clustering, while SCI presents the best latency performance. In \cite{127} it is presented a wide description of these technologies and Yeo {\it et al} \cite{100} present a benchmark-oriented study about the topic.\\

\subsection{Data management}
\label{subsec:Data_Management}

\noindent{Distributed Visualization deals with terabyte order datasets over heterogeneous platforms. The storage and utilization of this information have specific implications, specially the required physical space, the I/O tasks to be performed in suitable time and the applications' expected data format. Therefore, three efforts have emerged as leading initiatives to determine standards in scientific large volume data management: HDF \cite{83}, CDF \cite{133} and netCDF \cite{84}.}\\

\newpage
HDF, CDF and netCDF provide a platform-independent library via a high-level API. The stored data can be of any dimensionality and of several forms (numerical, string, images), it can be randomly read or written, unitarily or in blocks. HDF employs a more flexible data model (hierarchical) than netCDF and CDF (multidimensional array) and, according to Li {\it et al} \cite{84}, this flexibility comes at the cost of higher processing loads. The three formats are nearly equal referenced in the literature, they present equivalent features and performance. The choice for one of them should consider compatibility with the target system in hardware, software and development language.

\subsection{Parallel File Systems}
\label{subsec:Parallel_IO}

\noindent{In Distributed Visualization, it is possible to have tens of machines simultaneously accessing the same tera byte dataset. A single disk device cannot cope with these needs.  {\it Parallel file systems} were designed to deal with this problematic. These systems are designed on a client-server basis with multiple servers running a sort of I/O daemon. The parallel file system strips the files and store them across these servers. To retrieve information, the system reassembles a desired file and transmits it to the client. The whole process occurs automatically via calls to a user level library. Other functionalities like permission checking for file creation, open, close and removal are supported by an auxiliary manager process that handles meta data during the system operation.

Parallel file systems have to be robust and scalable, conform to existing I/O APIs for backward compatibility, maintain addressing file semantics, provide transaction support and be easy to use and install. Among the most popular implementations of parallel file systems for commodity PCs are the Lustre system \cite{144}, from Cluster File Systems Inc., released as open-source software, and the Parallel Virtual File System (PVFS) \cite{145}, also open-source, both for the Linux platform. The later is in its second release, which presents a number of improved features and a new design.

\newpage
Real parallel file systems are very complex. Maybe that is the reason why there is just a few implementations available for PC clusters. Comparing the options is not simple, as their complexity confer them a great number of features that are difficult to benchmark. Margo {\it et al} \cite{147} perform an extensive analysis of PVFS, Lustre and GPFS, however no categorical conclusions are drawn being up to the analyst to decide which one to choose. With the release of Lustre as open-source and with the emergence of PVFS version 2, these systems tend to evolve providing regular new features.

\subsection{Operating System}
\label{Operating_System}

\noindent{A report from Silicon Graphics observes \cite{94} that the operating system (OS) is replicated at each machine of a PC cluster leading to costs increase for each new node. License expenses, memory and processing requirements of each operating system instance sum up to a great burden. According to Yeo {\it et al} \cite{100}, besides these factors, the choice for the operating system in a cluster must consider: {\it manageability}, management and administration of local and remote resources; {\it stability}, robustness against system failures with system recovery; {\it performance}, optimized efficiency for OS tasks; {\it extensibility}, ability to easily integrate cluster-specific extensions; {\it scalability}, scale without performance degradation; {\it support}, user and system administrator support; and {\it heterogeneity}, support to multiple architectures to define a cluster consisting of heterogeneous hardware.}

Another consideration is the OS configuring likeness to enable variable configurations and customized optimizations. Choices in the market point to Unix proprietary solutions, to expensive Windows easy-to-set systems and to low-cost flexible (open-source) Linux systems. According to the worldwide top 500 hundred supercomputers list, reported by the Forbes magazine \cite{97}, the Linux platform has beaten competitors as the main choice for supercomputing.\\

\subsection{Technologies Summarization}
\label{subsec:Tech_Summarization}

\noindent{To link much of the information provided so far, in figure \ref{fig:Optimized_IO} we present the I/O structure of a cluster of computers with optimized data access. At the top layer is the {\it parallel execution} which is responsible for the data processing according to one of the parallelisms described in section \ref{Subsection_Algorithms}. These algorithms are load-balanced according to the discussion carried out in section \ref{Subsection_Load_Balancing}.}

\begin{figure}[ht]
		\centering		
\includegraphics[width=0.40\textwidth]{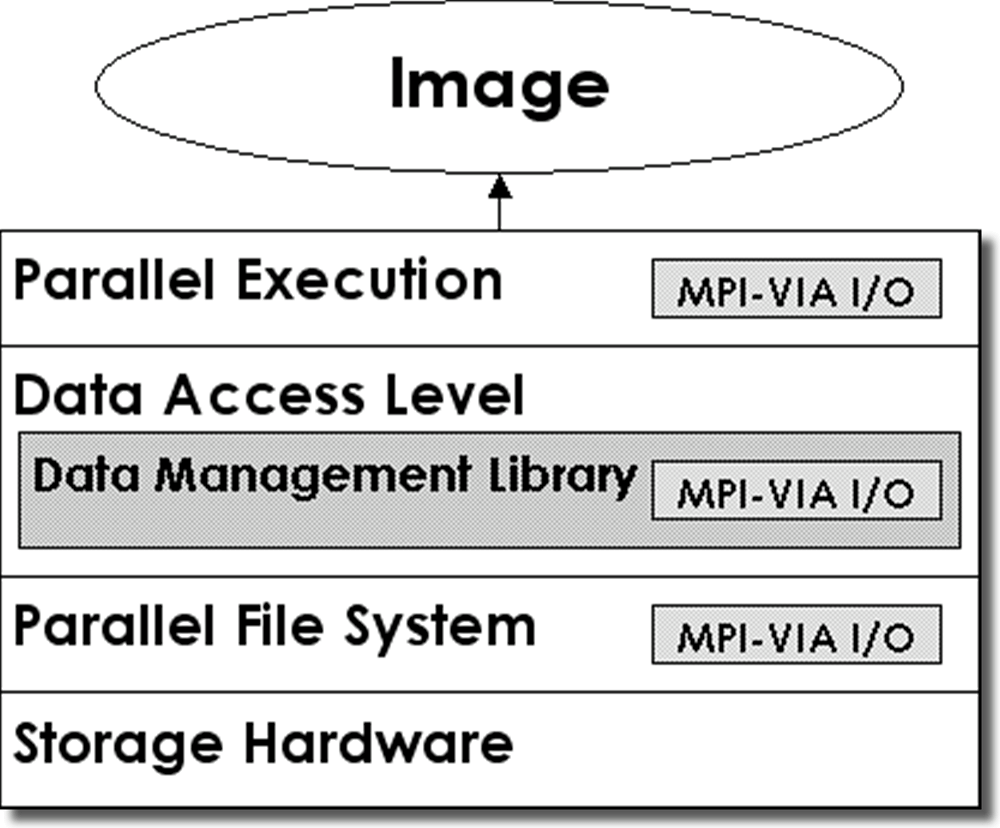}		
		\caption{The layers of a Distributed Visualization system. Storage devices are at the lowest layer abstracted by parallel file system access. The MPI standard, along with VIA technology, provides easy multi-point communication for data management libraries, such as HDF and netCDF, that feed the application with the data to be processed. At the highest layer, the parallel execution is performed to produce visualization images.}
	  \label{fig:Optimized_IO}		
\end{figure}

\indent{This model illustrates the state-of-the-art implementation strategy for Distributed Visualization concerning large datasets. The methodologies and technologies to be used at each layer depend on several factors as discussed along the text. Currently, works in the literature deal about finding the better settings for this model and/or to simplify it with more abstracting layers. This last issue is introduced in the next section in the form of distributed parallel rendering libraries.}
 
\indent{The layers in figure \ref{fig:Optimized_IO} are assisted by optimized network technologies, as presented in section \ref{Subsection_Network_Issues}. Among these technologies, the MPI standard, combined with VIA technology, is present at each layer of the model by providing simplified optimum access to remote data. To efficiently promote {\it data access}, it is necessary to use {\it data management} libraries, like those presented in section \ref{subsec:Data_Management}. These libraries access the information at the lowest layer, where lies the {\it storage hardware} providing voluminous data access. To abstract the storage hardware, {\it parallel file systems}, like the ones described in section \ref{subsec:Parallel_IO}, provide parallel high-performance transparent access.}

\section{Distributed parallel rendering libraries}
\label{Section_Distributed_Parallel}

\noindent{Attempts have been carried out to abstract Distributed Visualization through graphical libraries. The goal is to allow ordinary graphical library calls and have simplified management of distributed processing units as a visualization cluster. Earlier works met this goal but are not based on commodity hardware. Later proposals address commodity PCs.}\\

\noindent\textbf{WireGL}

\noindent{The WireGL library \cite{36} replaces the OpenGL driver to enable OpenGL in Distributed Visualization environments. By preserving the OpenGL API, applications can run on top of WireGL without recompilation and be provided with performance speedups, according to Humphreys {\it et al} \cite{36}. WireGL supports one or multiple clients simultaneously sending commands and data to one or multiple servers in sort-first parallelism. It intercepts regular OpenGL commands and send them to servers over the network. It also implements a network protocol for geometry communication and performs final image reassembly in software or hardware.}\\

\noindent\textbf{Chromium}

\begin{figure}[ht]
		\centering		
\includegraphics[width=\ProportionOfText\textwidth]{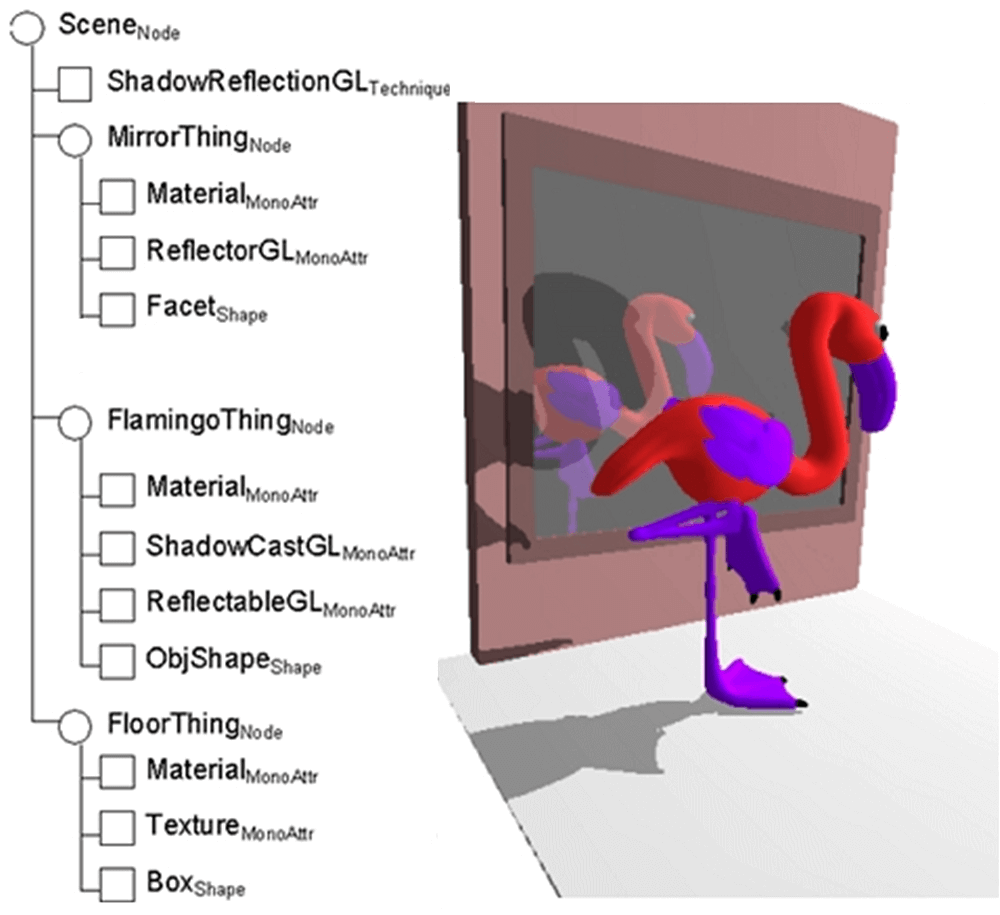}		
		\caption{Example of a scenegraph defined with the {\it Generalized Scene Graph API} \cite{107}. The hierarchical structure of the graph provides eased handling of complex scenes via propagated operations along the paths of the graph. Reproduced with permission granted by J{\"u}rgen D{\"o}llner and Klaus Hinrichs.}
	  \label{Scenegraph}		
\end{figure}

\noindent{Chromium \cite{37} is an advanced derivation from WireGL that similarly overlays OpenGL for compatibility. It supports the use of stream processing units, or SPUs, that perform specific rendering tasks. The SPUs can be chained to achieve a complex rendering execution. Chromium's architecture primes for its general orientation and flexibility. The SPU chain can be configured arbitrarily and both sort-first and sort-last parallelisms have been achieved, according to Humphreys {\it et al} \cite{37}. The drawback of Chromium's architecture is that its performance is influenced by its stream orientation, which cannot efficiently exploit frame coherence.}\\

\noindent\textbf{OpenSG}

\noindent{OpenSG \cite{69} is a scene graph multi-threaded API, specially designed for Virtual Reality applications. The scene graph metaphor (or hierarchical graphics database) organizes a graphical model as a graph that can manage visual entities hierarchically. By maintaining a copy of the scene graph for each thread, the threading system copes with distributed rendering because various servers can simultaneously respond to interaction (graph changes). To do so, OpenSG bears a client-server setup to replicate data on multiple machines that receive broadcasts informing of graph changes every frame. The library is flexible and can be used for implementing sort-first, sort-middle and sort-last algorithms. A similar library, also scene graph oriented, is the OpenRM project \cite{128}.}

The general orientation and high-level style of these libraries cause them to be less scalable than specific optimized implementations. This limitation is clearly demonstrated by Gribble {\it et al} \cite{149} who ported their Simian project both to a customized cluster implementation and to the top of the Chromium framework for performance comparison. Other issues are flexibility and compatibility.

\section{Conclusions}
\label{Section_Conclusions}

\noindent{We have surveyed basic concepts on Distributed Visualization. The presented content aims at elucidating what a Distributed Visualization system is, how it is characterized and what issues involve its design and implementation. We provide to beginners an introductory direction both for research and development and, for more experienced readers, we provide an analytical view of such systems. We have focused on distributed parallel rendering architectures, a cluster-based systematization that has popped up in the literature as works that explore flexible commodity low-cost PCs. These implementations have reached great performance levels and scalable architectures that evolve to the pace of market innovations.}
	
Many challenges still have to be bypassed in Distributed Visualization. Although the higher performance of PC clusters, their power is far from workstations as Silicon Graphics's InfiniteReality4 enabled systems, which scales up to 20.6 Gpixel textured antialiased pixels filled per second, or further. Robust real-time rendering for dynamic datasets is also an open challenge. Of-the-shelf Distributed Visualization software to amplify collaborative analytical work has not been accomplished either. We expect that this work can stimulate the quest for these goals by providing a source of information about the Distributed Visualization expertise.\\

\noindent{\textbf{Aknowledgements} Thanks to Funda\c{c}\~ao de Amparo \`{a} Pesquisa do Estado de S\~ao Paulo (FAPESP), Conselho Nacional de Desenvolvimento Cient\'ifico e Tecnol\'ogico (CNPq) and Coordena\c{c}\~ao de Aperfei\c{c}oamento de Pessoal de N\'ivel Superior (Capes).}

\begin{spacing}{0.9}
\bibliographystyle{plain}

\begin{thebibliography}{10}

\bibitem{127}
{IEEE} computer society task force on cluster computing - cluster computing
  white paper - version 2.0.
\newblock 2000.
\newblock 119 pages.

\bibitem{96}
Silicon graphics's white paper 3353 - infinitereality4 graphics - the ultimate
  in visual realism, 2002.

\bibitem{94}
Silicon graphics's white paper 3661 - choosing a visualization system for your
  organization, 2004.

\bibitem{104}
Dolphin interconnect solutions inc., 2005.
\newblock http://www.dolphinics.com/support/; accessed April, 2009.

\bibitem{91}
Infiniband trade association, 2005.
\newblock http://www.infinibandta.org; accessed April, 2009.

\bibitem{133}
Nasa - cdf home page, 2005.
\newblock http://cdf.gsfc.nasa.gov; accessed April, 2009.

\bibitem{83}
Ncsa hdf home page, 2005.
\newblock http://hdf.ncsa.uiuc.edu/; accessed April, 2009.

\bibitem{82}
The quake project, carnegie mellon university and san diego state university,
  2005.
\newblock http://www.cs.cmu.edu/\~{}quake/; accessed April, 2009.

\bibitem{98}
Top 500 supercomputer sites, top 500 statistics, 2009.
\newblock http://www.top500.org; accessed April, 2009.

\bibitem{130}
Baker, M., Farrell, P.~A., Ong, H., and Scott, S.~L.
\newblock Via communication performance on a gigabit ethernet cluster.
\newblock {\em Proceedings of the 7 th International Euro-Par Conference -
  Lecture Notes in Computer Science, vol. 2150/2001, Springer Verlag}, 2001.

\bibitem{128}
Bethel, E.~W.
\newblock White paper: Sort-first distributed memory parallel visualization and
  rendering with openrm scene graph and chromium.
\newblock 2003.
\newblock R3vis Corporation - 19 pages.

\bibitem{88}
Boden, N., Cohen, D., Felderman, R., Kulawik, A., Seitz, C.~L., Seizovic, J.,
  and Su, W.
\newblock Myrinet: A gigabit-persecond local area network.
\newblock {\em IEEE Micro}, 15(1):29--36, February 1995.

\bibitem{129}
Cameron, D. and Regnier, G.
\newblock {\em The Virtual Interface Architecture - 1st edition}.
\newblock Intel Press, 2002.
\newblock 210 pages.

\bibitem{145}
Carns, P.~H., III, W. B.~L., Ross, R.~B., and Thakur, R.
\newblock Pvfs: A parallel file system for linux clusters.
\newblock {\em Proceedings of the 4th Annual Linux Showcase and Conference},
  pages 317--327, 2000.

\bibitem{107}
D{\"o}llner, J. and Hinrichs, K.
\newblock A generalized scene graph api.
\newblock {\em Vision, Modeling, Visualization 2000 (VMV 2000) -
  Saarbr{\"u}cken. Akademische Verlagsgesellschaft.}, pages 247--254, 2000.

\bibitem{20}
Ellsworth, D.
\newblock {\em Polygon Rendering for Interactive Visualization on
  Multicomputers}.
\newblock Phd thesis, University of North Carolina, 1996.

\bibitem{111}
Ellsworth, D., Good, H., , and Tebbs, B.
\newblock Distributing display lists on a multicomputer.
\newblock {\em Computer Graphics}, 24(2):147--154, 1990.

\bibitem{99}
Farley, D.~L.
\newblock Performance comparison of mainframe, workstations, clusters, and
  desktop computers.
\newblock Technical Report NASA/TM-2005-213505, National Aeronautics and Space
  Administration (NASA), 2005.
\newblock 26 pages.

\bibitem{93}
Fernando, R.
\newblock Trends in gpu evolution.
\newblock {\em Annual Conference of the European Association for Computer
  Graphics - Industrial Seminars 2 Session}, 2004.
\newblock
  http://eg04.inrialpes.fr/Programme/IndustrialSem\\inar/PPT/Trends\_in\_GPU\_%
Evolution.pdf, Accessed April, 2009.

\bibitem{25}
Foley, J.~D., Dam, A.~v., Feiner, S.~K., and Hughes., J.~F.
\newblock {\em Computer Graphics: Principles and Practice}.
\newblock Addison-Wesley, 2 edition, 1997.

\bibitem{29}
Garcia, A. and Shen, H.-W.
\newblock An interleaved parallel volume renderer with pc-cluster.
\newblock In {\em Eurographics Workshop on Parallel Graphics and
  Visualization}, pages 51--59, Blaubeuren, Germany, 2002.

\bibitem{149}
Gribble, C., Cavin, X., Hartner, M., and Hansen, C.
\newblock Cluster-based interactive volume rendering with simian.
\newblock Technical Report TR UUCS-03-017, University of Utah, 2003.
\newblock 7 pages.

\bibitem{132}
Gropp, W., Lusk, E., Doss, N., and Skjellum, A.
\newblock A high-performance, portable implementation of the mpi message
  passing interface standard.
\newblock {\em Parallel Computing}, 22(6):789--828, 1996.

\bibitem{85}
Gropp, W., Lusk, E., and Thakur, R.
\newblock Using mpi-2: Advanced features of the message passing interface.
\newblock Technical report, MIT Press, 1999.
\newblock ISBN 0-262-57132-3, 382 pages.

\bibitem{32}
Haber, R.~B. and McNabb, D.~A.
\newblock Visualization idioms: A conceptual model for scientific visualization
  systems.
\newblock In Shriver, B., Neilson, G.~M., and Rosenblum, L., editors, {\em
  Visualization in Scientific Computing}, pages 74--93. IEEE, 1990.

\bibitem{36}
Humphreys, G. and al., P. H.~e.
\newblock Wiregl: A scalable graphics system for clusters.
\newblock In {\em Proceeding of SIGGRAPH}, pages 129--140, Los Angeles, CA,
  USA, 2001.

\bibitem{37}
Humphreys, G., Houston, M., Ng, R., Frank, R., Ahern, S., Kirchner, P.~D., and
  Klosowski, J.~T.
\newblock Chromium: A stream-processing framework for interactive rendering on
  clusters.
\newblock In {\em Proceedings of the 29th annual conference on Computer
  graphics and interactive techniques}, pages 693--702, San Antonio, Texas,
  USA, 2002.

\bibitem{144}
Inc., C. F.~S.
\newblock White paper - lustre: A scalable, high-performance file system.
\newblock 2002.
\newblock 13 pages.

\bibitem{131}
Kim, J.-S., Kim, K., and Jung, S.-I.
\newblock Building a high-performance communication layer over virtual
  interface architecture on linux clusters.
\newblock In {\em ICS '01: Proceedings of the 15th international conference on
  Supercomputing}, pages 335--347. ACM Press, 2001.

\bibitem{42}
Kurc, T.~M., Aykanat, C., and Ozguc, B.
\newblock A comparison of spatial subdivision algorithms for sort-first
  rendering.
\newblock {\em Lecture Notes in Computer Science}, 1225:137--146, 1997.

\bibitem{84}
Li, J., Liao, W.-K., Choudhary, A., Ross, R., Thakur, R., Gropp, W., Latham,
  R., Siegel, A., Gallagher, B., and Zingale, M.
\newblock Parallel netcdf: A high-performance scientific i/o interface.
\newblock In {\em Proceedings of Supercomputing 2003 Conference}, 2003.

\bibitem{90}
Liu, J., Chandrasekaran, B., Wu, J., Jiang, W., Kini, S., Yu, W., and Buntinas,
  D.
\newblock Performance comparison of mpi implementations over infiniband,
  myrinet and quadrics.
\newblock In {\em Proceedings of the ACM/IEEE SC2003 Conference}, Phoenix,
  Arizona, USA, 2003.
\newblock 58 pages.

\bibitem{44}
Lombeyda, S., Moll, L., Shand, M., Breen, D., and Heirich, A.
\newblock Scalable interactive volume rendering using off-the-shelf components.
\newblock In {\em IEEE Symposium on Parallel and Large-Data Visualization and
  Graphics}, pages 115--121, San Diego, CA, USA, 2001.

\bibitem{97}
Lyons, D.
\newblock Linux rules supercomputers.
\newblock {\em Forbes Magazine}, 2005.
\newblock http://www.forbes.com/2005/03/15/cz\_dl\_0315lin\\ux\_print.html;
  accessed April, 2009.

\bibitem{147}
Margo, M.~W., Kovatch, P.~A., Andrews, P., and Banister, B.
\newblock An analysis of state-of-the-art parallel file systems for linux.
\newblock Technical report, San Diego Supercomputer Center - University of
  California, 2004.
\newblock 28 pages.

\bibitem{50}
Molnar, S., Cox, M., Ellsworth, D., and Fuchs, H.
\newblock A sorting classification of parallel rendering.
\newblock {\em IEEE Computer Graphics and Applications}, 14(4):23--32, July
  1994.

\bibitem{51}
Montrym, J., Baum, D., Dignam, D., and Migdal, C.
\newblock Infinitereality: A real-time graphics system.
\newblock In {\em Proceedings of SIGGRAPH}, pages 293--302, 1997.

\bibitem{53}
Morel, K., Wylie, B., and Pavlakos, C.
\newblock Sort-last parallel rendering for viewing extremely large data sets on
  tile displays.
\newblock In {\em Proc. IEEE Symposium on Parallel and Large-Data Visualization
  and Graphics}, pages 85--92, San Diego, CA, USA, 2001.

\bibitem{54}
Mueller, C.~A.
\newblock {\em The Sort-First Architecture for Real-Time Image Generation}.
\newblock Phd thesis, The University of North Carolina at Chapel Hill, 2000.

\bibitem{110}
Networkworld.
\newblock Technology to tie together servers swells, 2004.
\newblock http://www.networkworld.com/news/2004/0412sp\\ecialfocus.html?page=1;
  accessed April, 2009.

\bibitem{89}
Petrini, F., Feng, W., Hoisie, A., Coll, S., and Frachtenberg, E.
\newblock The quadrics network: High-performance clustering technology.
\newblock {\em IEEE Micro}, 22(1):46--57, 2002.

\bibitem{61}
Roble, D.~R.
\newblock A load balanced parallel scanline z-buffer algorithm for the ipsc
  hypercube.
\newblock In {\em Proceedings of Pixim}, pages 177--192, Paris, France, 1988.

\bibitem{64}
Samanta, R., Funkhouser, T., and Li, K.
\newblock Parallel rendering with k-way replication.
\newblock In {\em Proceedings of the IEEE 2001 Symposium on Parallel and
  Large-data Visualization and Graphics}, pages 75--84, 2001.

\bibitem{68}
Sutherland, I., Sproull, R., and Schumacker, R.
\newblock A characterization of ten hidden surface algorithms.
\newblock {\em ACM Computing Surveys}, 6(1):1--55, March 1974.

\bibitem{135}
Thakur, R., Lusk, E., , and Gropp, W.
\newblock Users guide for romio: A high-performance, portable mpi-io
  implementation.
\newblock Technical Report Technical Memorandum ANL/MCS-TM-234, Mathematics and
  Computer Science Division, Argonne National Laboratory, 2004.
\newblock 15 pages.

\bibitem{69}
The-OpenSG-Staff.
\newblock Opensg starter guide.
\newblock http://www.opensg.org/downloads/OpenSG-1.2.0-UserStarter.pdf;
  accessed April, 2009.

\bibitem{12}
Touma, C. and Gotsman, C.
\newblock Triangle mesh compression.
\newblock {\em Graphics Interface}, pages 26--34, 1998.

\bibitem{71}
Upson, C. and al, e.
\newblock The application visualization system: A computational environment for
  scientific visualization.
\newblock In {\em IEEE Computer Graphics and Applications}, volume~9, pages
  30--42, 1989.

\bibitem{100}
Yeo, C.~S., Buyya, R., Pourreza, H., Eskicioglu, R., Graham, P., and Sommers,
  F.
\newblock Cluster computing: High-performance, high-availability, and
  high-throughput processing on a network of computers.
\newblock pages 521--551, 2005.

\bibitem{77}
Zhu, H., Chan, K.~Y., Wang, L., Cai, W., and See, S.
\newblock Dpbp: A sort-first parallel rendering algorithm for distributed
  rendering environments.
\newblock In {\em IEEE International Conference on Cyberworlds}, pages
  214--220, Marina Mandarin Hotel, Singapore, 2003.

\end{thebibliography}

\end{spacing}

\end{document}